# Estimating the proportion of differentially expressed genes in comparative DNA microarray experiments*

## Javier Cabrera[1] and Ching-Ray Yu[1]

*Rutgers University*

**Abstract:** DNA microarray experiments, a well-established experimental technique, aim at understanding the function of genes in some biological processes. One of the most common experiments in functional genomics research is to compare two groups of microarray data to determine which genes are differentially expressed. In this paper, we propose a methodology to estimate the proportion of differentially expressed genes in such experiments. We study the performance of our method in a simulation study where we compare it to other standard methods. Finally we compare the methods in real data from two toxicology experiments with mice.

## 1. Introduction

The human genome and a number of other genomes have been almost fully sequenced, but the functions of most genes are still unknown. The difficulty is that gene expression is only one of the pieces of cellular processes sometimes called biological pathways or networks, and it is not yet possible to observe these pathways directly. DNA microarray technology has made it possible to quantify and compare relative gene expression profiles across a series of conditions many thousands of genes at a time. By identifying groups of genes that are simultaneously expressed the guesswork of reconstructing biological pathways is expedited. The information collected through the years on genes that participate on biological pathways or networks has been used to construct GO (Gene Ontology Consortium [8]). The information on differentially expressed genes from a microarray experiment is contrasted with the groupings that are known according to existing GO and a determination is made on whether or not a certain cellular process is taking place. In addition there might be a few genes that are differentially expressed in the experiment but were not known to be part of the biological process. These genes become candidates for further extending the pathway and will be confirmed by further experimentation and also by searching for annotations that describe their function in other processes.

However, how to determine biological differentially expressed genes accurately is a nontrivial issue. Microarray experiments are high throughput in the sense that they evaluate the expression levels of thousands of genes at a time but with little replications. It is often the case that the number of replicate chips (biological, or technical) is 3 to 5 per condition. In addition the distributions of gene expressions across samples tend to be skewed and/or heavily tailed and hence they do not follow

---

a normal distribution. In this situation, permutation tests and traditional t-tests do not work very well because they have very low power.

One way to improve the power of the test is to incorporate the GO information to the process. Fisher's exact test (Fisher [7]) has been proposed as a way to detect if a particular subgroup of genes as a whole is differentially expressed. The test is applied to a two-way table of the indicator variable detecting the significance of the individual gene versus the indicator variable of the group. Another test is to consider the test statistic computed by Mean- Log-P, mean(-log(p-value)), (Pavlidis et al. [13] and Raghavan et al. [14]), of the genes in the group and compare this to the distribution of the statistics under a random subset of genes.

On the other hand, if when applying real data on GO, the number of differentially expressed genes overall is large, then the Fisher's exact test or Mean-Log-P test would still have low power. In order to overcome these problems we propose a new model approach, which consists of the following steps:

1. Estimate the proportion of differentially expressed genes.
2. Estimate the distribution of p-values for genes that are not differentially expressed. One would expect that this distribution is uniform but this is not the case in many examples that we have studied. The reason might be related to the processing of the data and the discarding of genes that take place at some stages of the process. Therefore the model has to estimate the distributions of null p-values by a semi-parametric or nonparametric method.
3. Estimate the distribution of p-values corresponding to differentially expressed genes.
4. Proceed by modeling the distribution of Mean-Log -P statistics for genes that belong to a subgroup or network. See Raghavan et al. [14], by using the estimators of steps 1-3.

In this paper we concentrate on step 1 of the procedure, which corresponds to the estimation of $\pi$. This quantity $\pi$ is important also in other situations, for example to calculate q-values (moderated p-values) proposed by Storey and Tibshirani [16]. For step 2-4 of the procedure, we will publish elsewhere as well. In Section 2 we propose a method and an algorithm for estimating $\pi$. In Section 3 we report the results of extensive simulation that support the performance of our method as well as comparison with other simpler methods.

Example *mice* and *mice*2: To illustrate the estimation of $\pi$, we apply our procedure for the mouse data sets from toxicology experiments (Amaratunga and Cabrera [3]). These datasets correspond to typical toxicology experiments where a group of mice is treated with a toxic compound and the objective is to find genes that are differentially expressed against samples from untreated mice.

*mice* and *mice*2 are two of the data sets that consist $n_1 = n_2 = 4$ mice in the control and treatment groups and total number of genes are $G = 4077$ from *mice* and $G = 3434$ for *mice*2 respectively. They represent two examples of cDNA chips, the first one *mice* has a high proportion $\pi$ of differentially expressed genes whereas *mice*2 has a much smaller $\pi$.

The data from such experiments consist of suitably normalized intensities: $X_{gij}$, where $g(g = 1, \ldots, G)$ indicates the genes on the microarray, $i(i = 1, 2)$ indexes the groups, and $j(j = 1, \ldots, n_i)$ is the $i$-th mouse in the $j$-th group. Our goal is to characterize $\Gamma$, a subset of genes, among the $G$ genes in the experiment that are differentially expressed across two groups.

Methods for determining $\Gamma$, researchers (e.g. Schena et al. [15]) use fold change, but they did not take variability into account. Subsequent improvements were t-test statistics (Efron et al. [6], Tusher et al. [17], and Broberg [5]), median-based methods (Amaratunga and Cabrera [1]) and Bayes and Empirical Bayes procedures (Lee et al. [10], Baldi and Long [4], Efron et al. [6], Newton et al. [12], and Lonnstedt and Speed [11]).

T-tests are the most widely used method for assessing differential expression. The assumption of the t-tests is that normalized intensities are approximately normally distributed with the same variance across the groups. i.e. $X_{gij} \sim N(\mu_{gi}, \sigma_g^2)$. For each gene g, a t-statistic is calculated in order to test null hypothesis $\mu_{g1} = \mu_{g2}$ and a p-value is generated. For small samples the t-test might be replaced by SAM or conditional t-test, Ct (Amaratunga and Cabrera [3]) in order to improve the power. Here we will follow the model proposed by Amaratunga and Cabrera [3] for the Ct method. Instead of trying to determine which genes are differentially expressed we will estimate the proportion of differentially expressed genes. Of course, as a consequence we could also produce an ordered list of genes that would be of interest to the biologist, but as we said above the entire procedure will be published elsewhere.

## 2. Statistical model and inference

The data for experiments typically consists of suitable iid normalized intensities:

$$(2.1) \qquad\qquad X_{gij} = \mu_g + \tau_{gi} + \sigma_g \epsilon_{gij},$$

where $\mu_g$ and $\sigma_g^2, g = 1, \ldots, G$, are the effect and variance of the $g$-th gene respectively, $\tau_{gi}$ is the effect of the $g$-th gene in the $i$-th group ($i = 1, 2$), and $j(j = 1, \ldots, n_i)$ indexes the samples. This is the same model in Amaratunga and Cabrera [3]. The treatment effect of the $g$-th gene is:

$$\tau_g = |\tau_{g2} - \tau_{g1}|$$

We assume that $\epsilon_{gij}$ are iid observations from an unknown distribution $F$ and we assume that $\sigma_g$ and $\tau_g$ are iid observations from unknown distributions $F_\sigma$ and $F_\tau$, respectively. $F_\sigma$ represents the distribution of the gene variances. $F_\tau$ is likely to have a mass at zero with probability $\pi$ representing the proportion of gene that are not differentially expressed. If the sample sizes were bigger the unknown distributions could be readily estimated by their respective cdf's but for small sample sizes the cdf's would produce very biased estimators. In the remainder of this section we will provide three procedures to estimate the three distributions $F$, $F_\tau$, and $F_\sigma$, which try to overcome the biases induced by small sample size.

In the model step:

1. Estimation of the error distribution $F_\epsilon$:

    In (2.1) when the number of samples per group is very small (3, 4, 5) and after residuals are subject to two constraints (sample mean $\bar{X} = 0$, sample standard deviation $s = 1$) then if we pool the residuals together, the empirical distribution that is obtained gives a very poor estimator of the error distribution $F$.

    For example: Suppose we sample 1000 genes from a normal distribution with two groups of subjects of sizes 4 and 4. The empirical distribution of the residuals is close to the true error distribution (which is standard normal) which is shown in the left-top graph of Figure 1, but if we also simulated the

t- distribution with df.=4, and 10 the qq-plot of the empirical distribution is not so good which is shown in the Figure 1.

One simple way to avoid this problem is to select a subset of genes $S_G$ that have small absolute t-values (say below 1 or some threshold that gives a large set of numbers). For each gene in $S_G$, both samples are pooled together and normalized by subtracting the gene mean and dividing by the standard deviation. If the sample size per group is very small (3, 4, 5) instead of the sample mean and standard deviation it is much better to use Huber M-estimator of location and scale (Huber [9]) as shown by Figure 1. This will result in a table of residuals $\hat{\epsilon}_{gij}$, $g \in S_G$. The error distribution $F_\epsilon$ is estimated by

$$(2.2) \qquad \hat{F}_\epsilon = EmpiricalCDF\{\hat{\epsilon}_{gij}, g \in S_G, i = 1, 2, j = 1, \ldots, n_i\},$$

Figure 1 shows the qq-plot for the estimated error distribution on t- distribution. The improvement is very clear.

2. Estimating $F_\sigma$:

We follow the method described in Amaratunga and Cabrera [2, 3]. They pointed out that the empirical distribution, $\hat{F}_\sigma$, of $s_g$ is a very poor estimator of the distribution $F_\sigma$, because on average $\hat{F}_\sigma$ is much more scattered than $F_\sigma$. They proposed an estimate $\tilde{F}_\sigma$ of $F_\sigma$ that shrinks $\hat{F}_\sigma$ towards its center and hence producing a better estimator of $F_\sigma$. A similar algorithm will be discussed in 3.

3. Estimating $F_\tau$: (determine the proportion of differential expressed genes)

We said earlier that $\tau_g$ is drawn from some distribution $F_\tau$. We expect that $F_\tau$ has a mass at zero of probability $F_\tau(0) \geq 0$, which represents the genes that are not differentially expressed. In order to estimate the probability $P(\tau_g = 0)$ we apply an algorithm that will produce an estimator $\tilde{F}_\tau$ such that the $E_{\tilde{F}_\tau}(\hat{F}_\tau^*(t)) = \hat{F}_\tau(t)$, where $\hat{F}_\tau^*(t)$ is the random variable representing the empirical cdf of $\tau^{**}$ at value $t$, which is constructed in following algorithm and $\hat{F}_\tau(t)$ represents the actual observed value.

The algorithm is as follows:

**Algorithm:**

Step 1:

**1.1)** Draw a random sample, $s^*$, from $\tilde{F}_\sigma$, which our estimate of the distribution of $\sigma$.

**1.2)** Estimate the error distribution $F_\epsilon$ with the empirical distribution $\hat{F}_\epsilon$ defined in (2.2).

**1.3)** Take a random sample (with replacement): $r_{gij} \sim \hat{F}_\epsilon$ for $i = 1, 2, j = 1, \ldots, n_i, g = 1, \ldots, N$.

**1.4)** Draw a sample $\tau_g^*$ from $\hat{F}_\tau(t) = I_{\{t \geq 0\}}$, where $I_{\{t \geq 0\}} = 1$ if $t \geq 0$ and $I_{\{t \geq 0\}} = 0$ if $t < 0$.

**1.5)** Construct the pseudo-data: $X_{g1j}^* = s_g * r_{g1j}, X_{g2j}^* = \tau_g^* + s_g * r_{g2j}$.

**1.6)** Reconstruct the distribution $F_{\hat{F}_\tau}^* = E(\hat{F}_\tau^* | \hat{F}_\tau)$, where $\hat{F}_\tau^*$ is the distribution of $\tau^{**}$ by pseudo-data: $\tau_g^{**} = |\bar{X}_{g2}^* - \bar{X}_{g1}^*|$.

**1.7)** Start by setting $\hat{F}_\tau^{(old)} = \hat{F}_\tau$.

**1.8)** Let $\hat{F}_\tau^{(new)} = \hat{F}_\tau(F_{\hat{F}_\tau^{(old)}}^{*-1}(\hat{F}_\tau))$.

**1.9)** Set $\hat{F}_\tau^{(old)} = \hat{F}_\tau^{(new)}$ and go to 1.3).

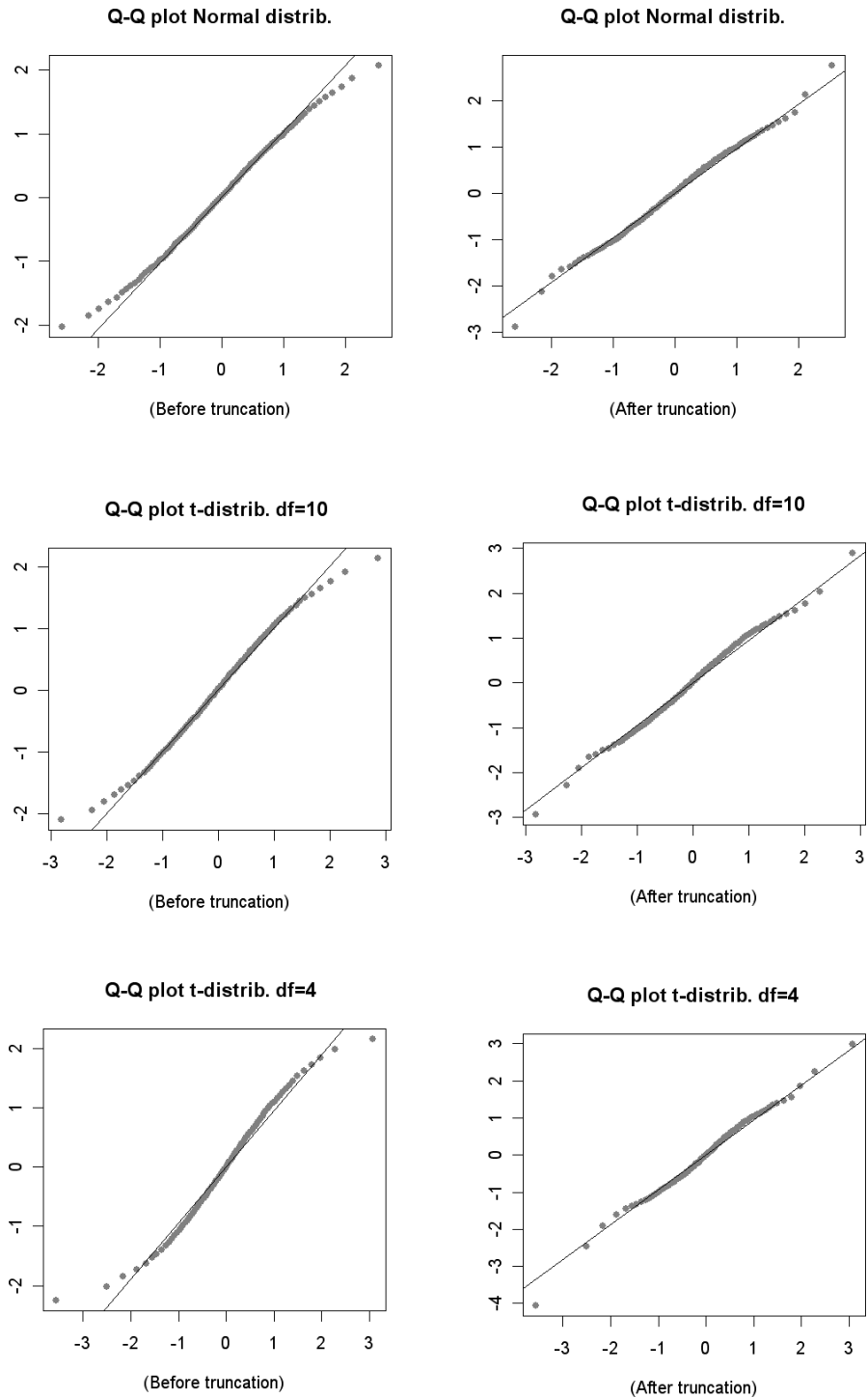

FIG 1. *A comparison of the error distribution estimates obtained from the empirical distribution (left) and our estimator (right), when the errors come from a Normal(0,1), $t_{10}$ and $t_4$ distributions.*

**1.10)** Iterate until convergence (approximately 100 iterations). At convergence we get our final estimate $\tilde{F}_\tau = \hat{F}_\tau^{(new)}$.

**1.11)** Give a cutoff point, say $\eta$, which is a 95% quantile of the final $\tilde{F}_\tau(t)$.

Step 2:

**2.1)** Repeat 1.4)-1.8) using all original data $X_{gij}$ and the estimated $\hat{F}_\tau$.

**2.2)** Get the estimated percentage of $\tau_g^{**}$ which is greater than $\eta \times 95\%$ quantile of standard normal.

**Theorem 2.1.** *At convergence the estimator $\tilde{F}_\tau$ is a fix point of the step in* **1.8)** *of the algorithm. That is $\tilde{F}_\tau = \hat{F}_\tau(F_{\hat{F}_\tau}^{*-1}(\tilde{F}_\tau))$, then we have*

$$(2.3) \qquad E_{\tilde{F}_\tau}(\hat{F}_\tau^*) = \hat{F}_\tau.$$

*Proof.* If the algorithm converges, then $\tilde{F}_\tau = \hat{F}_\tau(F_{\hat{F}_\tau}^{*-1}(\hat{F}_\tau))$. Thus

$$\hat{F}_\tau \circ \tilde{F}_\tau^{-1} \circ \hat{F}_\tau = F_{\hat{F}_\tau}^* = E(\tilde{F}_\tau | \tilde{F}_\tau) = \tilde{F}_\tau$$
$$\Rightarrow \hat{F}_\tau \circ \tilde{F}_\tau^{-1} = \tilde{F}_\tau \circ \hat{F}_\tau^{-1}$$
$$\Rightarrow (\hat{F}_\tau \circ \tilde{F}_\tau^{-1})^2 = I$$
$$\Rightarrow \hat{F}_\tau \circ \tilde{F}_\tau^{-1} = I$$
$$\text{or } \hat{F}_\tau \circ \tilde{F}_\tau^{-1} = -I \text{ (impossible, since } \hat{F}_\tau, \tilde{F}_\tau \geq 0)$$

$$E_{\tilde{F}_\tau}(\hat{F}_\tau^*) = E_{\tilde{F}_\tau}(\tilde{F}_\tau) = \tilde{F}_\tau = \hat{F}_\tau. \qquad \square$$

**Remark 1.** Base on our simulations, the algorithm converges in at most 100 iterations.

**Remark 2.** At convergence, $\tilde{F}_\tau$ is very close to $\hat{F}_\tau$ and $\hat{F}_\tau^*$ is also very close to $\tilde{F}_\tau$, such that we have nice result $E_{\hat{F}_\tau}(\hat{F}_\tau^*) = \hat{F}_\tau$.

**Remark 3.** This is a two-stage estimation method. We split data into two pieces. One is non-informative data, which produces a good estimation of the error distribution. The other is the informative data, we use shrinkage method to estimate the distribution of $\tau_g$, which gives the better result.

**Performance assessment:** To assess the performance of this method, we simulated data points, which are normally and independently distributed.

1. $X_{gij} \sim N(\tau_g, 1)$, where $G = 10000, n_1 = n_2 = 4$ and we assume that $G_{sig} = 1000, \ldots, 9000$ of $G$ genes were differentially expressed between two groups and their difference was $\delta$, i.e. $\tau_g = \delta(\delta = 1, 2)$ for all $g = 1, \ldots, G_{sig}$, and $\tau_g = 0$ otherwise.
2. $X_{gij} \sim N(\tau_g, \sigma_g^2)$, where $G = 10000, n_1 = n_2 = 4$ and we assume that $G_{sig} = 1000, \ldots, 9000$ of $G$ genes were differentially expressed between two groups and their difference was $\delta = 1, 2$, for all $g = 1, \ldots, G_{sig}$, and $\tau_g = 0$ otherwise and $\sigma_g^2$ are chi-square distributed with degrees of freedom 3. We calibrate the mean of $\sigma_g^2$ to 1. i.e. $\sigma_g^2/3$.

We compare our method to permutation tests and t-tests using a threshold of 0.05 to determine significance. These two methods are standard in biological applications. Our method is much more accurate than other two methods (Table 1-4, Figure 2). Each cell in the table is the mean (standard deviation) based on 10 times simulations on each condition. In Figure 2, the straight line represents the true values and the red line is obtained by the smooth spline function. We also calculate the pFDR of our method in different values of lambda (Table 5-6). pFDR decreases when the true value increases.

## 3. Discussion and extensions

In this paper we propose an algorithm for estimating the proportion of differentially expressed genes in a microarray experiment. We also show that the estimator of the distribution of the variance converges to a fix point. We performed a simulation study to check the performance of our estimate and it is shown to be "satisfactory" and we show that our method has better performance than other alternatives such as permutation tests and standard two-sample t- test. The simulations were performed under normal and gamma error distribution and with constant variances and chi-square variances. In addition we illustrate the method with real data examples on *mice* and *mice*2 (Table 7, Figure 3). In the real data examples we obtained estimates of the proportion of significant genes that were more realistic than those produced by the other methods. Hence, this algorithm gives us more accurate prediction to detect differential genes.

This same method is generally extendable to other more complicated modeling procedures such as the one-way ANOVA F-test and other linear models. The same model is used and the same ideas are easily extendable into a second paper. Another paper will deal with the GO issues, by modeling the p-values and getting a null distribution that will be used to detect differentially expressed gene network and subsets.

TABLE 1
*Normal(0,1)*

| $\delta$ | true $\lambda$ | 0.1 | 0.2 | 0.3 | 0.4 | 0.5 | 0.6 | 0.7 | 0.8 | 0.9 |
|---|---|---|---|---|---|---|---|---|---|---|
| 1 | t-test | 0.066 | 0.085 | 0.103 | 0.119 | 0.136 | 0.154 | 0.171 | 0.186 | 0.207 |
|   |   | (0.002) | (0.003) | (0.002) | (0.003) | (0.003) | (0.005) | (0.004) | (0.004) | (0.004) |
| 1 | Permutation | 0.039 | 0.051 | 0.063 | 0.073 | 0.085 | 0.096 | 0.107 | 0.116 | 0.130 |
|   | test | (0.002) | (0.002) | (0.002) | (0.003) | (0.003) | (0.003) | (0.003) | (0.003) | (0.003) |
| 1 | New method | 0.071 | 0.163 | 0.226 | 0.282 | 0.304 | 0.422 | 0.473 | 0.479 | 0.518 |
|   |   | (0.058) | (0.091) | (0.084) | (0.072) | (0.049) | (0.081) | (0.105) | (0.145) | (0.120) |
| 2 | t-test | 0.109 | 0.171 | 0.234 | 0.294 | 0.354 | 0.415 | 0.474 | 0.534 | 0.595 |
|   |   | (0.002) | (0.002) | (0.003) | (0.003) | (0.003) | (0.004) | (0.005) | (0.004) | (0.004) |
| 2 | Permutation | 0.074 | 0.120 | 0.168 | 0.214 | 0.259 | 0.305 | 0.350 | 0.397 | 0.442 |
|   | test | (0.003) | (0.002) | (0.002) | (0.003) | (0.004) | (0.003) | (0.005) | (0.005) | (0.004) |
| 2 | New method | 0.087 | 0.196 | 0.321 | 0.431 | 0.522 | 0.635 | 0.720 | 0.823 | 0.923 |
|   |   | (0.020) | (0.022) | (0.034) | (0.033) | (0.030) | (0.045) | (0.034) | (0.022) | (0.021) |

TABLE 2

$N(0,a), a \sim \chi^2_{(3)}/3$

| $\delta$ | true $\lambda$ | 0.1 | 0.2 | 0.3 | 0.4 | 0.5 | 0.6 | 0.7 | 0.8 | 0.9 |
|---|---|---|---|---|---|---|---|---|---|---|
| 1 | t-test | 0.066 | 0.087 | 0.110 | 0.134 | 0.157 | 0.180 | 0.204 | 0.227 | 0.252 |
| | | (0.001) | (0.004) | (0.002) | (0.004) | (0.003) | (0.004) | (0.003) | (0.004) | (0.003) |
| 1 | Permutation test | 0.045 | 0.060 | 0.077 | 0.095 | 0.112 | 0.129 | 0.148 | 0.163 | 0.182 |
| | | (0.002) | (0.002) | (0.002) | (0.003) | (0.002) | (0.003) | (0.004) | (0.003) | (0.002) |
| 1 | New method | 0.079 | 0.145 | 0.153 | 0.301 | 0.327 | 0.436 | 0.513 | 0.576 | 0.577 |
| | | (0.072) | (0.096) | (0.040) | (0.069) | (0.062) | (0.119) | (0.138) | (0.138) | (0.116) |
| 2 | t-test | 0.105 | 0.172 | 0.237 | 0.303 | 0.370 | 0.435 | 0.498 | 0.565 | 0.630 |
| | | (0.002) | (0.002) | (0.003) | (0.003) | (0.004) | (0.003) | (0.003) | (0.006) | (0.003) |
| 2 | Permutation test | 0.080 | 0.134 | 0.186 | 0.241 | 0.295 | 0.347 | 0.400 | 0.451 | 0.508 |
| | | (0.003) | (0.002) | (0.003) | (0.004) | (0.003) | (0.004) | (0.004) | (0.005) | (0.005) |
| 2 | New method | 0.111 | 0.207 | 0.311 | 0.413 | 0.514 | 0.609 | 0.712 | 0.811 | 0.914 |
| | | (0.027) | (0.034) | (0.032) | (0.030) | (0.025) | (0.022) | (0.017) | (0.018) | (0.015) |

TABLE 3

$Gamma(1, 1)$

| $\delta$ | true $\lambda$ | 0.1 | 0.2 | 0.3 | 0.4 | 0.5 | 0.6 | 0.7 | 0.8 | 0.9 |
|---|---|---|---|---|---|---|---|---|---|---|
| 1 | t-test | 0.067 | 0.094 | 0.123 | 0.150 | 0.178 | 0.207 | 0.233 | 0.264 | 0.292 |
| | | (0.002) | (0.004) | (0.003) | (0.004) | (0.004) | (0.004) | (0.004) | (0.004) | (0.003) |
| 1 | Permutation test | 0.053 | 0.075 | 0.099 | 0.120 | 0.143 | 0.168 | 0.190 | 0.213 | 0.237 |
| | | (0.001) | (0.002) | (0.003) | (0.003) | (0.004) | (0.003) | (0.003) | (0.006) | (0.003) |
| 1 | New method | 0.059 | 0.151 | 0.225 | 0.310 | 0.321 | 0.377 | 0.482 | 0.504 | 0.626 |
| | | (0.043) | (0.035) | (0.075) | (0.062) | (0.099) | (0.110) | (0.094) | (0.119) | (0.107) |
| 2 | t-test | 0.108 | 0.177 | 0.246 | 0.313 | 0.381 | 0.450 | 0.521 | 0.588 | 0.657 |
| | | (0.002) | (0.002) | (0.003) | (0.003) | (0.005) | (0.003) | (0.004) | (0.005) | (0.005) |
| 2 | Permutation test | 0.090 | 0.151 | 0.212 | 0.272 | 0.330 | 0.391 | 0.454 | 0.514 | 0.576 |
| | | (0.003) | (0.002) | (0.002) | (0.003) | (0.004) | (0.004) | (0.003) | (0.005) | (0.004) |
| 2 | New method | 0.126 | 0.232 | 0.310 | 0.417 | 0.515 | 0.613 | 0.712 | 0.802 | 0.912 |
| | | (0.048) | (0.045) | (0.024) | (0.020) | (0.023) | (0.010) | (0.015) | (0.014) | (0.013) |

TABLE 4

$t_5$

| $\delta$ | true $\lambda$ | 0.1 | 0.2 | 0.3 | 0.4 | 0.5 | 0.6 | 0.7 | 0.8 | 0.9 |
|---|---|---|---|---|---|---|---|---|---|---|
| 1 | t-test | 0.065 | 0.086 | 0.109 | 0.130 | 0.153 | 0.174 | 0.197 | 0.218 | 0.243 |
| | | (0.003) | (0.003) | (0.005) | (0.004) | (0.003) | (0.004) | (0.003) | (0.004) | (0.002) |
| 1 | Permutation test | 0.043 | 0.058 | 0.075 | 0.090 | 0.106 | 0.122 | 0.138 | 0.153 | 0.170 |
| | | (0.002) | (0.002) | (0.004) | (0.003) | (0.003) | (0.003) | (0.004) | (0.003) | (0.003) |
| 1 | New method | 0.074 | 0.141 | 0.208 | 0.212 | 0.319 | 0.368 | 0.490 | 0.530 | 0.641 |
| | | (0.060) | (0.100) | (0.065) | (0.074) | (0.080) | (0.091) | (0.133) | (0.128) | (0.084) |
| 2 | t-test | 0.112 | 0.177 | 0.241 | 0.309 | 0.373 | 0.440 | 0.507 | 0.575 | 0.639 |
| | | (0.002) | (0.002) | (0.002) | (0.005) | (0.003) | (0.003) | (0.004) | (0.005) | (0.005) |
| 2 | Permutation test | 0.083 | 0.136 | 0.190 | 0.246 | 0.298 | 0.352 | 0.408 | 0.461 | 0.517 |
| | | (0.002) | (0.003) | (0.002) | (0.004) | (0.003) | (0.004) | (0.004) | (0.006) | (0.006) |
| 2 | New method | 0.113 | 0.205 | 0.309 | 0.411 | 0.516 | 0.610 | 0.718 | 0.811 | 0.918 |
| | | (0.030) | (0.013) | (0.028) | (0.027) | (0.022) | (0.016) | (0.027) | (0.017) | (0.013) |

TABLE 5

*pFDR for our method with Normal(0,1) error distribution*

| true $\lambda$ | 0.1 | 0.2 | 0.3 | 0.4 | 0.5 | 0.6 | 0.7 | 0.8 | 0.9 |
|---|---|---|---|---|---|---|---|---|---|
| $\delta = 1$ | 0.5471 | 0.3768 | 0.3823 | 0.2372 | 0.2372 | 0.1924 | 0.1486 | 0.0860 | 0.0482 |
| | (0.1679) | (0.1371) | (0.0537) | (0.0535) | (0.0535) | (0.0354) | (0.0363) | (0.0209) | (0.0131) |
| $\delta = 2$ | 0.1963 | 0.1924 | 0.2416 | 0.1533 | 0.1215 | 0.0965 | 0.0841 | 0.0601 | 0.0465 |
| | (0.0753) | (0.0741) | (0.0876) | (0.0393) | (0.0406) | (0.0242) | (0.0255) | (0.0093) | (0.0112) |

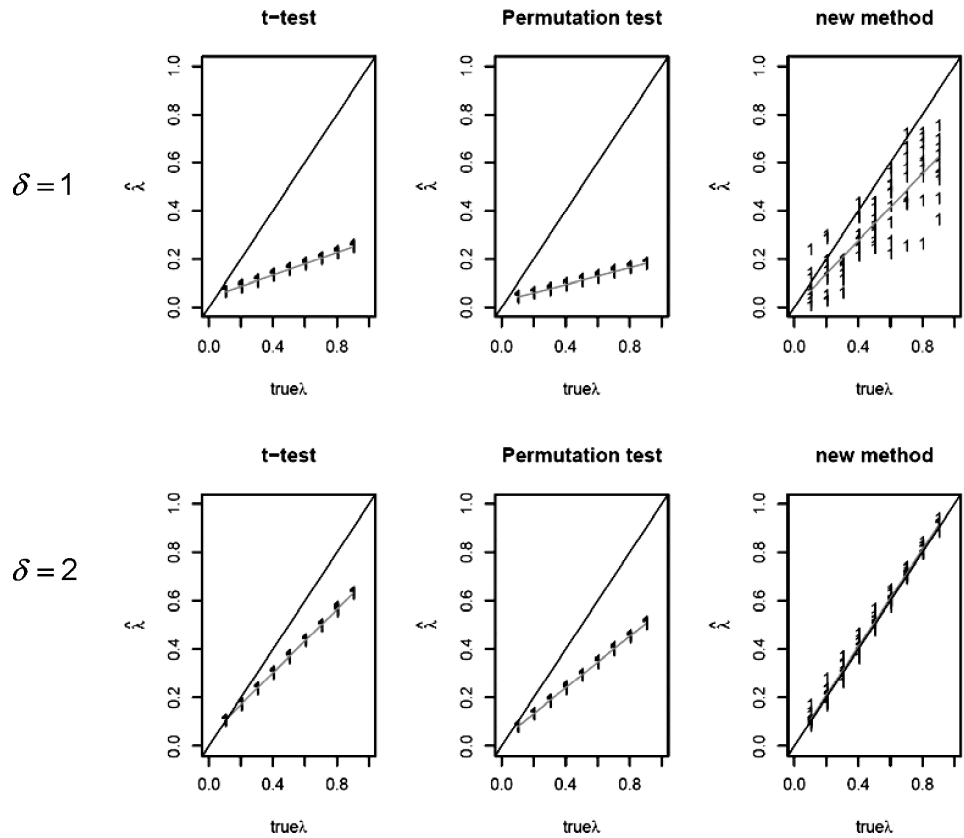

FIG 2.  *Example comparing our method to the Permutation and t methods. The true errors are* $N(0, \sigma^2), \sigma^2 \sim \chi^2_{(3)}/3$.

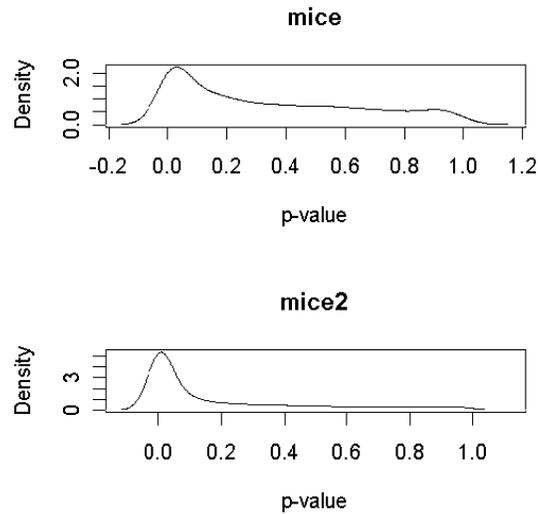

FIG 3.  *Density estimators for the p-values obtained from two toxicology datasets.*

TABLE 6
*pFDR for our method with $Normal(0, \sigma^2), \sigma \sim \chi_{(3)}^2/3$ error distribution*

| true $\lambda$ | 0.1 | 0.2 | 0.3 | 0.4 | 0.5 | 0.6 | 0.7 | 0.8 | 0.9 |
|---|---|---|---|---|---|---|---|---|---|
| $\delta = 1$ | 0.634 | 0.480 | 0.375 | 0.323 | 0.233 | 0.185 | 0.102 | 0.094 | 0.047 |
| | (0.069) | (0.060) | (0.060) | (0.040) | (0.053) | (0.048) | (0.018) | (0.017) | (0.0135) |
| $\delta = 2$ | 0.325 | 0.226 | 0.167 | 0.139 | 0.119 | 0.107 | 0.074 | 0.063 | 0.037 |
| | (0.099) | (0.054) | (0.042) | (0.022) | (0.020) | (0.016) | (0.017) | (0.014) | (0.0047) |

TABLE 7
*Results for the three methods applied to
two real examples from toxicology*

| Estimated $\pi$ | *Mice* | *Mice2* |
|---|---|---|
| *t − test* | 0.245 | 0.499 |
| *Permutation test* | 0.220 | 0.443 |
| *New method* | 0.107 | 0.363 |